\documentclass[12pt,preprint]{aastex}
\usepackage{amssymb,amsmath,float}

\begin{document}

\title{\large\bf Transient Heliosheath Modulation}

\author{J J Quenby$\,^{1}$ and W R Webber$\,^{2}$}

\affil{\(^{\scriptstyle 1} \){Blackett Laboratory, Imperial College,
London, SW7~2BZ, UK.}\\}

\affil{\(^{\scriptstyle 2} \){Department of Astronomy, New Mexico
State University, Las Cruces, USA}\\}

\begin{abstract}

Voyager 1 has explored the solar wind-interstellar medium interaction region
between the Terminal Shock and Heliopause, following the intensity distribution
of galactic cosmic ray protons above 200 MeV energy. Before this component
reached the expcted galactic flux level at 121.7 AU from the sun, four episodes
of rapid intensity change occured with a behaviour similar to that found in 
Forbush Decreases in the inner solar system, rather than that expected from a 
mechanism related to models for the Long Term Modulation found closer to the
sun. Because the mean solar wind flow is both expected and observed to be 
perpendicular to the radial direction close to the Heliopause, an explanation
is suggested in terms of transient radial flows related to possible heliopause
boundary flapping. It is necessary that the radial flows are at the sound speed
found for conditions downstream of the Terminal Shock and that the relevant 
cosmic ray diffusion perpendicular to the mean field is controlled by 'slab'
fluctuations accounting for 20 \% or less of the total power in the field
variance. However, additional radial drift motion related to possible
north to south gradients in the magnetic field may allow the inclusion of
some diffusion according to the  predictions of a theory based upon the
presence of 2-D turbulence 
The required field gradients may arise
due to field variation in the field carried by solar plasma flow deflected 
away from the solar equatorial plane.
Modulation amounting to a total 30 \% drop
in galactic intensity requires explanation by a combination of transient
 effects. 

\end{abstract}

\section{INTRODUCTION}

 Since the provision of early models of heliosheath modulation (Potgieter
 and le Roux 1989, Quenby et al. 1990), it has been assumed that this
 region between the terminal shock and the heliopause is a location where
 a substantial fraction of the solar modulation of the galactic cosmic ray
 intensity occurs (see review by Potgieter (2008)). A very recent description 
of heliosheath modulation in a spherically symmetric approximation is due to 
Webber et al. (2013b) while (Potgieter (2013) povides a recent general
review. The
 heliopause represents the boundary beyond which the interstellar cosmic ray
intensity would be encountered. Strauss et al. (2013) 
 have recently questioned this assumption and mention
 various possibilities of increased particle scattering
beyond the heliopause as the interstellar
 field wraps around the Heliopause.
Models and observations related to the interaction of the interstellar medium
(ISM) with the heliosphere allow a solar wind terminal shock 
and a heliosheath lying between this shock and the heliopause. These models
 suggest that the heliosheath 
comprises a low latitude region where the magnetic structure is determined 
by reconnection of the sector structure fields
and a high latitude region where field lines connect back to
the solar wind (Opher el al. 2012). The field is carried by the solar wind
 as it is diverted to high latitudes and back downwind of the interstellar
flow. No interstellar bow shock is expected 
(McComas et al. 2012) but the external field pressure causes asymmetry in
the terminal shock (Opher et al. 2006) and may also explain
 intermitent observation
 of shock accelerated cosmic rays (Jokipii et al. 2004, Stone et al. 2005)
 prior to the termination shock crossing.\\
 The dramatic Voyager 1 observation of two sudden increases of the greater 
 than 200 MeV proton galactic cosmic ray (GCR) component near the heliopause 
 reached at 121.7 AU ( Webber and McDonald, 2013a, Stone et al., 2013)
 presents a challenge to current ideas of modulation within the heliosheath. 
 It is the purpose of this paper to provide a simple class of models for these
 sudden increases in terms analagous to the cause
of Forbush Decreases in the inner solar system. There should exist transient,
enhanced radial plasma flows which may or not be accompanied by changes
 in magnetic field gradients
suitable to yield enhaced particle drift speeds.
A companion paper (Webber and
Quenby, 2014) discusses the observations of the two extra-ordinary increases
in GCR intensity in relation to magnetic field data and the force field 
modulation model.\\ 
         
\section{VOYAGER DATA}

The starting point of this work lies in data obtained by Webber and McDonald
  (2013a)
from the Voyager 1 CRS instrument (Stone et al. 1977). We concentrate on the 
period of the final increase in the $>$ 200 MeV proton intensity to
attain the expected galactic flux value, witnessed by the steady intensity 
distribution beyond 121.7 AU. The increase around 2012 day 240 is represented
by a change of scaled count rate from 4.08 to 4.58 in 3.7 days  
(see figure 1, Webber and McDonald, 2013a). In figure 2 of Webber and Quenby 
(2014) it is seen that these data correspond to the final, step like
 GCR intensity increase before a galactic value is reached and that it is
 preceeded by two fluctuations of similar magnitude over the previous month.   
These two preceeding flutuations do not reach the galactic GCR value. They 
could be regarded as two Forbush Decreases exhibiting rapid recovery
of the flux to pre-decrease levels or above. If this 
structure in the GCR intensity is convected past Voyager 1 with velocity 
$V_{100}$ in units of 100 km/sec, the spatial gradient is typically 
$0.56/V_{100}$
per AU. Since the spacecraft is moving at only 0.07 AU per week, it is very
unlikely that the intensity structure is stationary in time. The regions of
observed high spatial gradient extend back to about 120.5 AU to include the
second region of high spatial gradient discussed by Webber and Quenby (2014).
A significant observation by these last authors lies in the relative modulation
of the 200 MeV protons and of 10 MeV galactic electrons. Both species exhibit
the step intensity increases but the relative modulation is higher 
by about a factor two for the electron channel \\
In order to estimate the likely mean solar wind velocity close to the
heliopause, we appeal to the models provided by Opher et al.(2012) who
provide alternatives, based upon whether or not reconnection in the sector
region is included. Some verification is provided by using the Low Energy
Charged Particle (LECP) Voyager 1 results to deduce the radial and tangential
flow components in the absence of a working plasma probe. Unfortunately the
numerical results need scaling from a modelled heliopause at 162 AU. 
The Opher et al. (2012) flow vectors are given, both at 120 AU which is 
within 2 AU of the heliopause and at 110 AU for the radial, tangential and
normal components 
in Table 1 for alternative models, either ignoring the low-latitude  sector 
structure or taking into account reconnection associated with the sector 
structure. \\

\begin{table}
\caption{Opher et al. (2012) Models and Voyager 1 Flow data 
in the Heliosheath in km/sec}
\label{symbols}
\begin{tabular}{@{}cccc}
\hline
Source & $V_{R}(120)$ & $V_{T}(120) $& $V_{N}(120)$\\ 
 \hline
 No-Sector & 44 & -30 & 68         \\
 Sector    &  4 &  -5 & 30       \\
 Data      &    &     &   \\
\hline
& $V_{R}(110)$ & $V_{T}(110)$ & $V_{N}(110)$ \\
\hline 
No-Sector & 52 & -71 & 50        \\
 Sector    & 12 & - 5 & 15      \\
 Data      & 22 & -36 & (29)\\
\hline
\end{tabular} 
\end{table}
The LECP keV data is given at 110 AU for two components. It appears that 
the experiment 
seems to agree better with the sector model for $V_{R}$ and the
 non-sector model
for $V_{T}$. However, the data exhibits very large fluctuations throughout the 
heliosheath. Negligible LEPC data is available for the normal, 
$V_{N}$ component which clearly becomes the dominant steady flow at 
these distances. We estimate the measured $V_{N}(110)$ as the mean ot the other
two components. Extrapolation of the experimental results suggests that
$V_{R}(120),V_{T}(120)\leq \pm 20$ km/sec.\\    
Magnetometer data obtained by Burlaga et al. (2013) during the heliopause 
crossing enable us to estimate the power in the fluctuations of the magnitude
of $\bf{B}$. These authors observe a period of outward pointing polarity 
in 2012 from DOY 150 to DOY 171, inward polarity from DOY 176 to DOY 202
and outward polarity from DOY 204 DOY 238, the last period including two
intervals where post-heliopause conditions are apparently encountered. 
Using data from figure 2 of
Burlaga et al. (2013), we give in Table 2 the mean and standard deviations of
 the fields in these 3 periods, neglecting the post-heliopause data. The
quoted experimental errors are $\pm $ 0.01 nT, so are not expected to greatly
distort the estimates of the standard deviation.\\

\begin{table}
\caption{Voyager 1 Field Magnitude and Standard Deviation, sd  }
\label{symbols}
\begin{tabular}{@{}cccc}
\hline
 Period       & DOY 150-171     & DOY 176-202  & DOY 204-238 \\
                      
\hline
 B, nT   & 0.182          & 0.339    & 0.225         \\
 sd, nT   &0.0690           & 0.0596   &  0.0735      \\
\hline
\end{tabular} 
\end{table}

An estimate of the correlation length of the field magnitude fluctuations may
be obtained from the power spectrum provided by Burlaga and Ness (2010) where
a break at $1.8 \times 10^{-7}$ Hz is found in the Voyager 1 data at 110 AU. If
the field is convected past the spacecraft at a relative speed of 46 km/sec,
as suggested by the data, the correlation length is 1.7 AU, taking into account
the Voyager 1 velocity\\      
Webber and Quenby (2014) show large changes in the magnetic field energy 
density well correlated with the large fluctuations in GCR intensity during
 the period day 208 to day 240. This observation provides indirect evidence
of significant change in plasma velocity.
Quenby and Webber (2013) have suggested that transient radial plasma
flow velocities near the boundary, triggered by large changes in plasma
pressure, could reach the sound speed obtained by Borovikov et al. (2011)
in simulations of post terminal shock conditions. Their value is 228 km/sec.
The attainment of this speed
out to near the heliopause depends on the plasma temperature not decaying
significantly. If the Alfven speed becomes the dominant fast mode speed, a 
simple estimate can be made based on the continuity of mass flow,
$nVr^{2}=$ constant where n is plasma number density at radial distance r. 
Using the observed field and the post-shock and non-sector velocities of 
previously mentioned models, we find an Alfven speed of 31 km/sec at 120 AU.\\
\section{ DIFFUSION MEAN FREE PATHS}

To carry out our calculation, we need
 to estimate possible values 
of the diffusion coefficients parallel and perpendicular to the mean field.
We will use a theoretical formulation which has achieved reasonable agreement
 with experimental results in the inner Heliosphere, but employing field data
obtained far out in the Heliosheath. The waves in the field model are composed
 80\% of a 2-dimensional
 component with fluctuation vectors perpendicular to both the mean field
and wave propagation direction and 20\% of a slab component with fluctuations
 perpendicular to the mean field but with wave propagation along the mean field.
As derived by Le Roux et al. (1999), this composite field model yields a 
parallel diffusion mean free path
\begin{equation}
\lambda_{||}=2.433 \frac{B^{2}}{b_{x,sl}^{2}}(\frac{P}{cB})^{1/3}
\lambda_{sl}^{2/3}\times F
\end{equation}
 where $P$ is particle rigidity, $B$ is the mean field, $b_{x,sl}$ is the 
x component of the slab field fluctuations, $\lambda_{sl}$ is the correlation 
length of the fluctuations, which are
assumed to be valid for all components and F is function
very close to unity in the present application. Pei et al. (2010) employ a 
very similar result to model the parallel diffusion coefficient throughout
the Heliosphere.\\
For the perpendicular diffusion coefficient, we follow the Le Roux et al. 
(1999) non-perturbative approach where
\begin{equation}
D_{sl}=\frac{1}{2}\lambda_{sl}A_{sl}^{2} ;~~ D_{2D}=\lambda_{2D}A_{2D}
\end{equation}
$D_{sl}$ describes the magnetic field wandering due to slab turbulence
with $A=b_{x,sl}/B$ with a corresponding correlation length $\lambda_{sl}$. 
$D_{2D}$ denotes the magnetic field line wandering for 2-D fluctuations.
$A_{2D}$ is the amplitude for this turbulence and 
$\lambda_{2D}$ is the corresponding
correlation length perpendicular to background field.\\
We note that
if the contribution of the 2-D fluctuations is ignored, the perendicular
diffusion coefficient becomes the expression given in Jokipii (1971)   
\begin{equation}
K_{\perp}=\frac{v}{4}\frac{P_{xx}(0)}{B^{2}}
\end{equation}
where $P_{xx}(0)$ is the power in one slab component at zero frequency.
If however, 2D turbulence dominates, we find from le Roux et al. (1999)
the modified value 
\begin{equation}
K_{\perp,2D}=\frac{1}{2}v\lambda_{2D}A_{2D}
\end{equation}

Le Roux et al. (1999) also consider the quasi-linear model of Chuvilgin and
Ptuskin (1993) whih also takes into account resonant and nonresonant
interactions. These last authors derive
\begin{equation}
K_{\perp}=0.5A^{2}K_{||}
\end{equation}
where A is the total fractional deviation in the field. We neglect a
relatively unimportant adjustment
to the numerical value given in this last equation to obtain agreement with 
some numerical simulations performed at 1 AU.
  
In the following section it will become apparent that the simple, Jokipii,
(1971) expression is more likely to satisfy the proposed modulation model.
  In Table 3, values of parallel diffusion
coefficients and mean free paths are calculated from the data of Table 2,
using the Le Roux et al. (1999) expression, equation (1),
 while perpendicular diffusion coefficients
and mean free paths are obtained from the same data using the Jokipii (1971)
expression. To estimate the power in the slab component, ie that due to 
wave propagation along the mean field direction, 
it is assumed that the dominant
fluctuation power is transverse. To first order,  
\begin{equation}
\frac{sd}{B}=\frac{1}{2}\frac{\delta B_{\perp}^{2}}{B^{2}}
\end{equation} 
and if only 20\% of the standard deviation results from the slab-like 
fluctuations, then
\begin{equation}
P_{xx}(0)=0.2\frac{\lambda_{sl}}{2\pi}\frac{ \delta B_{\perp}^{2}}{2}
\end{equation}
It is assumed that $\lambda_{sl}=1.7$ AU.  
\begin{table}
\caption{Parallel and Perpendicular Diffusion Coefficients and Mean Free Paths
derived from Table 2 Data for 200 MeV protons }
\label{symbols}
\begin{tabular}{@{}cccc}
\hline
 Period                 & DOY 150-171     & DOY 176-202  & DOY 204-238 \\
 \hline
$K_{||} AU^{2}s^{-1}$ & $7.42\times10^{-3}$ & $1.3\times10^{-2}$ & $8.01\times10^{-3}$         \\
$\lambda_{||}$ AU  & 19.6           & 34.3   &  21.2      \\
$K_{\perp} AU^{2}s^{-1}$ & $2.9\times10^{-6}$ & $1.35\times10^{-6}$ &$ 2.52\times10^{-6}$\\
$\lambda_{\perp} AU$ & $7.68\times10^{-3}$ & $3.57\times10^{-3}$ &$ 6,64\times10^{-3}$\\
\hline
\end{tabular} 
\end{table}
Note that Pei et al (2010) find values of $\lambda_{\perp}\sim 10^{-2}$ AU for 
100 MeV protons far out in the heliosphere based on models for the development
of turbulence.\\
Using the 2D turbulence model as expressed by Le Roux et al. (1999), we obtain
$K_{\perp,2D}=4.5\times10^{-4}$ AU$^{2}$/s or $\lambda_{\perp,2D}=1.2$ AU for the 
period DOY 176-202 of Table 3. The Chuvilgin and Ptuskin (1993) expression
yields $K_{\perp,2D}=1.2\times10^{-3}$ AU$^{2}$/s or $\lambda_{\perp,2D}=3.0$ AU\\
\section{A TRANSIENT 3-D MODULATION APPROXIMATION}

In the first model of
the approximation of this section the aim is to attempt to explain the
 observed large radial gradients near the heliopause in terms of
transient radial velocities and diffusion parameters which seem to be possible
according to the limited plasma information available.
The argument could be reversed to
make use of the cosmic ray data to suggest nesessary values for the plasma 
parameters. No attempt will be made to provide a 3-D modulation solution
 encompassing a large Heliospherical volume. Instead, a Cartesisn geometry is 
adopted to describe a limited region close to the heliopause, rather than 
spherical polar coordinates which are better as a basis for overall 
Heliospheric modulation modelling. A quasi-steady state will be assumed,
as has been applied previously to the modelling of short-term changes in the 
inner Heliosphere where field structure, convected past the observer, carries
a particle distribution which only changes slowly in time. (eg Mulligan et al.
2009 and Quenby et al. 2008)  
Setting ${\bf x}$ in the radial direction
${\bf y}$ in the azimuthal direction and ${\bf z} $ to complete the right
handed set, the Fokker-Planck equation in the Sun's reference frame
for the differential number density $U({\bf r},t)$ is (eg Quenby (1984),
\begin{equation}
\frac{\partial U}{\partial t} + {\bf \nabla}{\bf S}+
\frac{\partial}{\partial T}(\frac{dT}{dt})U=0
\end{equation}
where the streaming is
\begin{equation}
{\bf S}=C{\bf V}U-\underline{K}.\nabla U
\end{equation}
and the Sun frame kinetic energy, T, loss rate is
\begin{equation}
\frac{dT}{dt}={\bf V}.\nabla(\frac{\alpha TU}{3})
\end{equation}
where $\alpha=(T+2m_{\circ})/(T+m_{\circ})$. $m_{\circ}$ is rest mass and the
Compton Getting factor C is
\begin{equation} 
C=(1-\frac{1}{3U}\frac{\partial}{\partial T} \alpha TU)
\end{equation}
The slope of the energy spectrum and $\alpha$ are assumed to be constant
 over the limited region of application of the above Fokker Planck equation,
thus both $C$ and the pressure term in the expression for
kinetic energy T are independent of position. The spectral slope at 200 MeV is 
obtained from Stone et al. (2013). Resolving components, we notice that
 perpendicular
 diffusion applies in the ${\bf x}$ 
and ${\bf z}$ directions while parallel diffusion applies in the ${\bf y}$
direction. We also allow the transverse diffusion coefficient to yield 
a drift 
velocity $V_{d,z}$ in the direction perpendicular to the plane 
of the 
spacecrafts orbit due to the large scale Parker field structure and a possible 
drift $V_{d,x}$ due to a field gradient 
in the ${\bf z} $ direction arising from lack of symmetry about the
equatorial plane.
The steady state Fokker-Planck becomes
\begin{eqnarray}
 K_{\perp}\frac{\partial^{2}U}{\partial x^{2}}-
(1.15V_{x}+V_{d,x})\frac{\partial U}
{\partial x}+K_{||}\frac{\partial^{2}U}{\partial y^{2}}-
1.15V_{y}\frac{\partial U}{\partial y} \nonumber  \\
+K_{\perp}\frac{\partial^{2}U}{\partial z^{2}}
-(1.15V_{z}+V_{d,z})\frac{\partial U}{\partial z} =0
\end{eqnarray}
Neglecting spatial variation in $ {\bf V}$ and $\underline K$ and specifying
$U_{\circ}$ as the differential number density at the origin of coordinates, 
that is at the bottom of 
the region of sudden intensity increase, a seperable, trial solution 
results in 
\begin{equation}
 U=U_{\circ}exp(\frac{(1.15V_{x}+V_{d,x})x}{K_{\perp}}+\frac{1.15V_{y}y}{K_{||}}+
\frac{(1.15V_{z}+V_{d,z})z}{K_{\perp}})
\end{equation}
Hence the solution tracks the initial, outward rise in intensity. $U_{\circ}$ is
determined by the boundary conditions on the heliopause and
the relative ease of entry of GCR along the three spatial directions. However
this work is not attempting a complete 3D solution, especially as the plasma
parameters along the three directions are poorly known. Rather we point out
the expected result that a local radial, $ {\bf x}$, directed gradient requires
a matching radial, outward directed plasma flow. In the absence of a suitable
 drift velocity, such a flow does not exist,
close to the heliopause
in the time independent Heliosphere model of Opher et al. (2012)
 Therefor,
radial gradients close to this boundary may require transient radial flows. 
We first explore a 'drift free' model. To 
obtain modulation below the GCR level near the heliopause, sudden 
bursts of radial plasma
velocity need to appear sufficiently frequently so there is always
one event within about 1 AU of the boundary if the series of  
steep depressions shown in Figure 1 of
Webber and McDonald, (2013a) are to be explained. It is within this region that
the intensity switches from near 
the GCR level to a modulated level. Any lesser 
occurence frequency would mean a further encroachment of the GCR level within
the heliosheath.\\  
As source of the transient radial plasma flows, we 
 postulate that the radial plasma flow 
corresponds to the flows required by Quenby and Webber (2013) who suggested
that the heliopause flapped with the speed of the fast mode, here identified as
the sound speed in the hot solar wind plasma. The flapping could be caused by 
pressure imbalance with the external medium as the solar wind plasma pressure
varies.\\    
In order to see if the diffusion and flow parameters discussed in sections 2 
and 3 can satisfy the observed gradient employing the transient modulation 
model, we find the radial cosmic ray
gradient from the adopted solution as given by
\begin{equation}
K_{\perp}\frac{\partial U}{\partial x}=1.15V_{x}+V_{d,x}
\end{equation}
On DOY 240, 2012, $\frac{\partial U}{\partial t_{s}} =0.032$ per day is
the rate of intensity change in the reference frame of V1. In the adopted
model DOY 240 is regarded as corresponding to the end of a field structure 
convected past V1 as the heliopause expands. The gradient then refers to the 
entry of particles into the region of reduced intensity by diffusion in
competition with the wind outward sweeping.
 Since the
measurement of the actual spatial gradient is determined by the adopted radial
wind speed, the above equation can be written as 
\begin{equation}
K_{\perp}=1.39\times10^{-6}V_{x,100}^{2}+1.19\times10^{-6}V_{d,x,100}.V_{x,100}
~~{\rm AU}^{2}{\rm s}^{-1}
\end{equation}
where $V_{x,100}$ and $V_{d,x,100}$ are respectively the radial and drift
velocities in units of 100 km/sec. 
Neglecting the addition drift term in $V_{100,d,x}$ 
on the right hand side of the above equation,  if the 
radial flow is $\sim$ 100 km/sec, the estimated values of $K_{\perp}$ found
in Table 3 seem to satisfy the measured gradient. Suitable flow values are
the sound speed estimated from post terminal shock conditions or the Opher
et al (2012) model which ignores the sector structure. However, this second,
model estimate does not provide transient effects. Models for $K_{\perp}$
which add 2-D turbulence onto the slab turbulence field line wandering 
so as to yield the dominant effect are 
less likely to satisfy the cosmic ray data.\\
The average drift motion at the epoch of observation is northward for positive
charged particles in the 
northern solar hemisphere because the average solar field is inward at the time
The drift magnitde $V_{d}$ from guiding centre theory is
\begin{equation}
V_{d}=\frac{1}{\omega R}\frac{2}{3}v^{2}=\frac{2vPc}{3BR}
\end{equation}
where $\omega$ is the cyclotron frequency, $P$ the rigidity
 and $R$ estimates both the field
line radius of curvature and the fractional field gradient. 
Using the mean measured field at 120 AU and assuming R=120 AU, the northward
 drift speed $V_{d,z}=54$ km/sec. 
However, transient changes show
both a field line polarity switch and a field magnitude gradient which is
 positive in the outward direction (Burlaga et al., 2013), thus 
reinforcing the northward proton drift. Hence
modulation can be enhanced in the northward direction, as compared with the
solar wind sweeping.\\
For charge e, the field gradient and curvature drift are aligned in the
 $\frac{1}{e} {\bf B x \nabla B}$ 
direction. Hence for protons,
 an episode of outward directed north hemisphere ${\bf B}$ together
 with a limited region of ${\bf \nabla B }$ directed in the $-{\bf z}$ 
direction produces an outward, ${\bf x }$ directed drift.
According to the
Opher et al. (2012) model with the sector structure included, gradients 
$(\nabla B)/B \approx 0.018$/AU may exist in the southward or $-{\bf z}$ 
direction on the edge of the sector region.
This would lead to a radial drift of 115 km/sec, outward for outward north
hemisphere field epochs. Neglecting for the moment the evidence provided by
the electron modulation, one may develop the idea of a drift dominated flow.
Burlaga et al. (2013) observed an
unexpected field increase close to the heliopause. Averaging over a time
period equivalent to a distance scale of 6 cyclotron radii for a 200 MeV 
proton, a field gradient $(\delta B)/B \approx 0.78$/AU
is obtained. If this also
 occured in the $-{\bf z}$ direction, perhaps as a transient effect, 
 a radial drift of 4900 km/sec would be possible.  The effect of adding these
various values of outward radial drifts on the required value of $K_{\perp}$
can be seen by employing  equation (15). As before, it is assumed that the
end of the passage of a field structure on DOY 240 is being considered.
Additionaly a typical plasma motion $\sim$100 km/s is used in order to 
estimate the spatial gradient. We find 
$K_{\perp}=6.3\times10^{-5}$ AU$^{2}$s$^{-1}$ is allowed if the field 
gradient is 0.78/AU directed southward. However, from the equation for the drift
speed, it is found that electrons of 10 MV rigidity drift two orders of 
magnitude slower than 200 MeV protons and are therefore relatively  
insensitive to field gradient effects. Since the electron modulation is 
greater than that of protons, it is very unlikely that drift can dominate the
modulation  near the heliopause. This is because the electron scattering 
mean free path at 10 MV is unlikely to show a large reduction compared with
that of a 200 MeV proton.   
In any case, there still seems to be 
shortfall by a factor near 10 as compared with the estimated 2D turbulence
model for perpendicular diffusion. It is perhaps significant that two 
periods of large radial gradient seen by Webber and McDonald (2013a)
centred on DOY 128 and 209 are in periods of outward pointing field. \\
Turning to the GCR sudden increase starting on day 128, the lack of detailed
correlation with change in magnetic field amplitude led Webber and Quenby 
(2014) to suggest the cause lies in structure $\sim$ 1 AU in extent. On the 
basis of the 
model provided here, the necessary sudden, radial increase in solar
wing plasma velocity, together perhaps with an
  outward, radial particle drift speed, must have 
occured between V1 and the heliopause and not at the location of V1. In fact,
since the major plasma flow is expected to be northwards out of the ecliptic
plane this close to the heliopause, there is no necessity for V1 to see the 
plasma changes responsible for the sudden increase 
being convected past the 
spacecraft prior to the event. For our model to explain the day 128 increase,
we have two requirements. The first is that the northward plasma flow brings
a large plasma pressure increase causing rapid outward expansion at 
the heliopause with or without a suitable $-{\bf z}$ directed field
 gradient to allow
enough outward GCR motion to counter inward diffusion. The second 
requirement is that the scale size of the changes to the plasma parameters
close or at the heliopause provide a depletion region of GCR intensity
where particle leakage from the $ \bf{y,z}$ directions is slower than
the current due to the outward convection or drift velocities. If our 
model is reasonable, the ocurrence
of the day 128 increase is both evidence for dramatic movement at or within
1 AU of the heliopause and of a scale size of the movement in directions 
parallel to the heliopause of several AU.

\section{MEAN FREE PATH RIGIDITY DEPENDENCE}
Webber and Quenby (2014) discuss the rigidity dependence of the modulation
during the period of the two increases, days 128 to 238. A problem    
arising from the data analysis reported by these authors
which the work presented here has not explained is the lack of apparent 
charge splitting in the relative modulation suffered by protons and alpha
particles. This splitting was first predicted by the force field 
approximation of Gleeson and Axford (1968). However, unlike the previous 
authors, we do not depend either on assuming no net streaming in a particular
direction or on integrating the solution between the spacecraft and the 
boundary of modulation. Instead, we simply consider the local values of the 
intensity gradients during the increases. This however limits the discussion
to the relative changes in electron and proton intensity, these being the
species for which we have detailed information.\\
From equation (13), the ratio electron to proton gradients in the x direction
if $V_{x}>>V_{d,x}$ is
\begin{equation}
\frac{1}{U_{\circ}(P_{e})}\frac{\partial U(P_{e})}{\partial x}/
\frac{1}{U_{\circ}(P_{p})}\frac{\partial U(P_{p})}{\partial x}
=\frac{K_{\perp,p}(P_{p})}{K_{\perp,e(P_{e})}}=
\frac{\beta_{p}\lambda_{\perp,p}(P_{p})}{\beta_{e}\lambda_{\perp,e}(P_{e})}
\end{equation}
where $P_{e}$, $P_{p}$ and $\beta_{e}$, $\beta_{p}$
 are the magnetic rigidities and velocities  
of the electron and protons observed. From Fig 2 Webber and Quenby, (2014),
the ratio of the fractional electron to proton intensity change
  measured at the step change around day 210, 2012 is 2.6.
Taking the electron mean energy as 10 MeV and the protons as corresponding to 
200 MeV, a mean free path dependence $\lambda_{\perp}\propto P^{0.4}$
is required to fit the diffusion of both species if the day 210 change 
represents the ratio of electron to proton intensity gradients. For the 
day 128 step, the gradient ratio is 1.6 and the exponent of the rigidity
dependence is 0.29.
. This fit favours the rigidity dependence of equation (5) if equation (1)
is used to define $K_{||}$. However, the absolute value of the perpendicular
diffusion coefficient is still more than two orders of magnitude too large to 
balance the expected transient convective radial flow.\\
A drift dominated model seems excluded by the relative size of the electron 
to proton step changes and therefore cannot be used to discuss the mean free
path rigidity dependence.  
  As a conclusion to this section (5) it seems that none of the expressions for
the perpendicular diffusion coefficient discussed simultaneously satisfy 
both the absolute magnitude and rigidity depndence implied on either of the two
transient modulation models suggested. A small addition to the Jokipii (1971),
equation (3) expression, dependent on a $\lambda_{||}\propto P^{1/3}$ due to
nonresonant interactions would seem to provide the closest overall fit. 
     

\section{CONCLUSIONS}
In the attempt to explain the unexpectedly large radial cosmic ray gradients 
observed around 120 AU an appeal has been made to the possible existence of
transient outward plasma flows, approaching the sound speed in the Heliosheath.
Interpreting the available magnetometer data to yield a suitable value
of the perpendicular diffusion coefficient, it was necessary to employ
a model of scattering dependent only on 20\% of the measured fluctuation power.
This power would constitute the percentage present in waves propagating along 
the magnetic field direction. 
The resulting modulation model is similar to
that expected from a succession of Forbush Decreases in the inner Heliosphere.
Field gradient 
particle drift effects can also play a role in modulation near the
heliopasuse. If these are sufficiently large,
additional diffusion based upon 2-D turbulant fields may 
explain the rigidity dependence of the modulation.
Transient changes in radial plasma speed or field gradient necessary to cause
the observed sudden GCR intensity changes do not necessarilly need to be located at the spacecraft but may alternatively occur between the spacecraft and the
heliopause. However, although we believe it worthwhile discussing the two
 transient modulation models
we provide, problems arise in satisfying the observed
rigidity dependence of the modulation in terms of available models 
and data relating to diffusion in the Heliosheath.      
Further information on the relevent plasma parameters,
especially transient effects, at 120 AU is required
to determine whether either of the proposed models will remain 
successful. Also
further ellucidation of the relation of magnetic field turbulence to the
actual cosmic ray diffusion perpendicular to the mean field seems required.\\

\section{ACKNOWLEDGMENTS}

W R Webber thanks his Voyager 1 colleagues from the CRS instrument,
 especially the late Frank McDonald and Project PI Ed Stone.
Magnetometer data were obtained by Len Burlaga and Norman Ness.
                    
\section{REFERENCES}
\noindent
 Borovikov S. N., Pogorelov N. V., Burlaga L. F., Richardson \\
\indent 
J. D., 2011, ApJ, 728 L21\\
\noindent
Burlaga L. F., Ness N. F., Acuna M. H., Richardson J. D.,\\
\indent
 Stone E. C., McDonald F. B.,2009 ApJ, 692, 1135 \\    
\noindent
 Burlaga l. F., Ness N. F., 2010, ApJ., 725, 1306\\ 
\noindent
Burlaga L. F., Ness N. F., 2012, ApJ., 749, 13\\ 
\noindent
 Burlaga L. F., Ness N. F., Stone E. C., 2013, \\
\indent
Science, 341, 150\\
\noindent
Chuvilgin L, G., Ptuskin V. S., 1993,\\
\indent
Astron. Astrophys., 279, 273\\
\noindent
Gleeson L.I., Axford W.I., 1968, Ap.J., 154, 1011\\
\noindent
Gurnett D. A., Kurth W. S., Cairns I. H., Mitchell J., 2006\\
\indent
in Heerikhuisen J.,Florinski V., Zank G. P., Pogorelov\\
\indent
N. V., eds, Physics of the Inner Heliosheath. American\\
\indent
Inst. Phys. p.129 \\
\noindent
Jokipii J.R., 1971, Rev. Geophys. Space Sci, 9, 27\\
\noindent
Jokipii J. R., Giacaloni J., Kota J., 2004 ApJ 611, L141\\
\noindent
Krimigis S. M.,
Roelof E. C., Decker R. B., Hill M. E., \\
\indent
2011, Nature 474, 359 \\
Krimigis S. M., Decker R. B., Roelof E. C.\\ 
\indent
Hill M. E., Armstrong T. P.,Gloeckler G.,\\
\indent
Hamilton D. C., Lanzerotti, L. J., 2013\\ 
\indent
Science, 341, 144\\   
\noindent
Le Roux J. A., Zank G. P., Ptuskin V. S., 1999,\\
\indent
J. Geophys. Res., 104, 24845\\ 
\noindent
McComas D.J., Alexashov D. A., Browski M., Fahr H.,\\
\indent
 Heerikhuisen J.,Izmodenov V., 
Lee M.A., Mobius\\
\noindent
Mulligan T., Blake, J. D., Shaul D., Quenby J. J.,\\
\indent
 Leske R. A., Mewaldt D. M., Galametz M., 2009,\\
\indent
 J Geophys. Res. Space Phys., 114, 13783\\
\indent
M. A., Pogorelov N., Schwadron N.A., Zank G.P., 2012\\
\indent
Science, 336, 1291\\  
 Nkosi,G. S., Potgieter  M. S., Webber W. R., 2011,\\
\indent
 Adv.in Space Res., 48, 1480\\
\noindent
Opher M., Stone E. C.,
Liewer P.C., 2006 ApJ. 640 L71 \\
\noindent
Opher, M., 
Drake J. F., Velli M., Decker R.B., Tooth G.,\\
\indent
2012, ApJ, 751, 80\\
Pei C., Bieber J. W., Breech B., Burger R. A.,\\
\indent
Clem J., Matthaeus W. H., 2010, \\
\indent
J. Geophys. Res.,115,A03103 \\
\noindent 
Potgieter M. S., Le Roux J. A., 1989, A \& A, 209, 406 \\ 
\noindent
 Potgieter, M. S., 2008, Advances in Space Sci. 41, 245\\
\noindent
 Potgieter, M. S., 2013, Living Reviews in \\
\indent
Solar Physics DOI:10.12942/lrsp-2013-3\\
\noindent
Quenby J. J., 1984, Space Sciences Review, 37, 201\\
\noindent
Quenby J. J., Lockwood J. A., Webber W. R., 1990,\\
\indent
 ApJ, 365, 365\\
\noindent
Quenby J. J., Mulligan T., Blake J. B.,\\
\indent 
Mazur J. G., Shaul D., 2008, J Geophys. Res., \\
\indent
Space Phys., 113, 12849\\
\noindent
Quenby J. J., Webber W. R., 2013, Monthly Notices \\
\indent
Royal Astronomical Society, 436, 3306\\ 
\noindent
Scherer K., Fichtner H., Strauss R.D.,\\ 
\indent
Ferreira S. E., Busning L., Potgieter M.S.,\\
\indent
2008, Ap J., 680, L105 \\
\noindent
\noindent
Stone E. C.,
Vogt R. E., McDonald F. B., Teegarden B. J.,\\
\indent
 Trainor J. F., Jokippi, J. R., Webber W. R., 1977,\\
\indent
 Space Science Reviews, 21, 355\\    
\noindent
Stone E, C., 
Cummings A. C., McDonald F. B., Heikkila\\
\indent
B., Lal N., Webber W. R., 2013,
Science, 341, 150  \\
Strauss R.D., Potgieter M.S., Ferreira S. E. S.,\\ 
\indent
Fichtner H., Scherer K., 2013,\\
\indent
 ApJ. Lett. 76 L18\\
\noindent
\noindent
 Webber W.R., McDonald F.B., 2013a,\\ 
\indent
Geophys.Res.Lett., 40, 1665\\
\noindent
 Webber W.R., Higbie P. R., McDonald F.B., 2013b,\\
\indent
 arxiv..org ;1308.1895\\
\noindent
Webber W.R., Quenby J J.,2014 Submitted for publication\\ 
\end{document}